\documentclass[twocolumn,aps]{revtex4-1}

\pdfpageattr {/Group << /S /Transparency /I true /CS /DeviceRGB>>}

\usepackage{graphicx} 
\usepackage{overpic}
\usepackage{enumitem}
\usepackage{amssymb}
\usepackage{courier}

\newcommand{\ket}[1]{\left|#1\right\rangle}

\begin{document}

\title{Low overhead quantum computation using lattice surgery}

\author{Austin G. Fowler, Craig Gidney}
\affiliation{Google Inc., Santa Barbara, California 93117, USA}

\date{\today}

\begin{abstract}
When calculating the overhead of a quantum algorithm made fault-tolerant using the surface code, many previous works have used defects and braids for logical qubit storage and state distillation. In this work, we show that lattice surgery reduces the storage overhead by over a factor of 4, and the distillation overhead by nearly a factor of 5, making it possible to run algorithms with $10^8$ T gates using only $3.7\times 10^5$ physical qubits capable of executing gates with error $p\sim 10^{-3}$. These numbers strongly suggest that defects and braids in the surface code should be deprecated in favor of lattice surgery.
\end{abstract}

\maketitle

\section{Introduction}

In \cite{Babb18}, we contributed overhead calculations based on the surface code \cite{Brav98,Denn02} using defects and braiding \cite{Raus07,Raus07d,Fowl12f} showing that non-trivial post-classical computations could be performed in hours using of order a million superconducting qubits. Here, we shall focus on lattice surgery, systematically building up the techniques required to implement an arbitrary algorithm and calculate the precise time and space overhead. We find that general algorithms can be implemented with nearly a factor of 5 fewer qubits and comparable time using lattice surgery. The results here were prepared in parallel with and are consistent with the recent work \cite{Liti18b}.

\section{Storage}

\begin{figure}
\begin{center}
\resizebox{55mm}{!}{\includegraphics{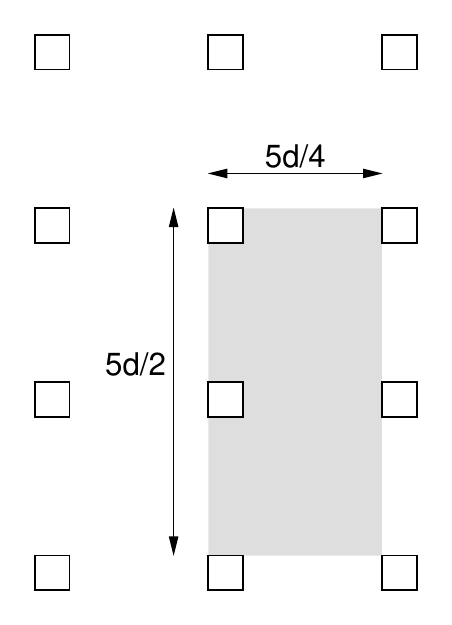}}
\end{center}
\caption{In terms of the code distance $d$, a double-defect logical qubit occupies $25d^2/8$ space to leading order. Each unit of $d$ represents two qubits, a data and measure qubit, so to leading order $12.5d^2$ physical qubits are required.}\label{double_defect}
\end{figure}

A double-defect logical qubit (Fig.~\ref{double_defect}) consists of two holes in a large non-rotated surface code such that each hole's circumference is no less than the code distance $d$, and the holes are also separated by at least $d$. In a simple regular packing, the kind of packing that has been used in many resource estimates, $12.5d^2$ physical qubits are required per logical qubit.

\begin{figure}
\begin{center}
\resizebox{65mm}{!}{\includegraphics{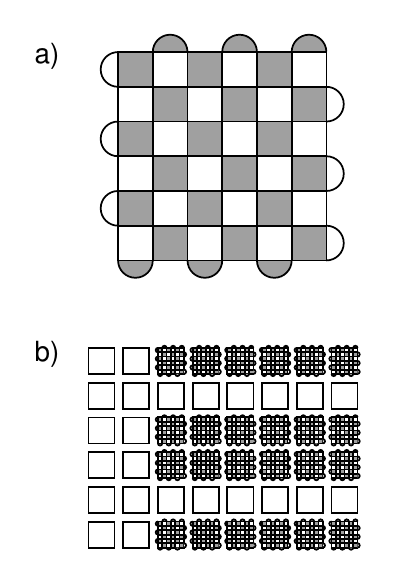}}
\end{center}
\caption{a) Distance $d=7$ rotated logical qubit. Dark regions represent X stabilizers, light regions represent Z stabilizers. Each region is associated with a measurement qubit. A data qubit is located at each intersection point of dark lines. b) Proposed layout of rotated logical qubits permitting local operations in parallel and easy movement of collections of logical qubits to the workspace on the left where multi-logical-qubit operations can take place.\label{rotated}}
\begin{center}
\resizebox{85mm}{!}{\includegraphics{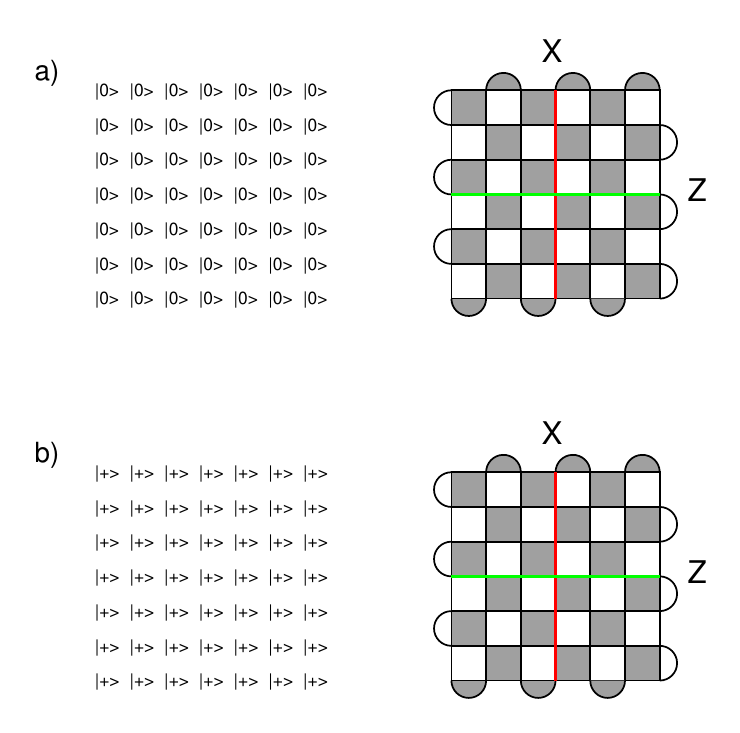}}
\end{center}
\caption{a) Preparation of $\ket{0_L}$. All data qubits are initialized to $\ket{0}$ then all stabilizers measured the code distance $d$ times. b) Preparation of $\ket{+_L}$. All data qubits are initialized to $\ket{+}$ then all stabilizers measured the code distance $d$ times.\label{initialization}}
\end{figure}

Lattice surgery \cite{Hors11}, makes use of rotated logical qubits (Fig.~\ref{rotated}a) consisting of $d^2$ data qubits and $d^2-1$ measure qubits. In this work, we shall arrange the rotated logical qubits being used in an algorithm in patterns similar to Fig.~\ref{rotated}b. Such arrangements enable logical qubits to be both acted on locally and easily moved to a shared workspace if required for multi-logical-qubit operations. To leading order, $3d^2$ physical qubits are required per logical qubit, better than a factor of 4 overhead saving.

\section{Logical initialization and measurement}

\begin{figure}
\begin{center}
\resizebox{75mm}{!}{\includegraphics{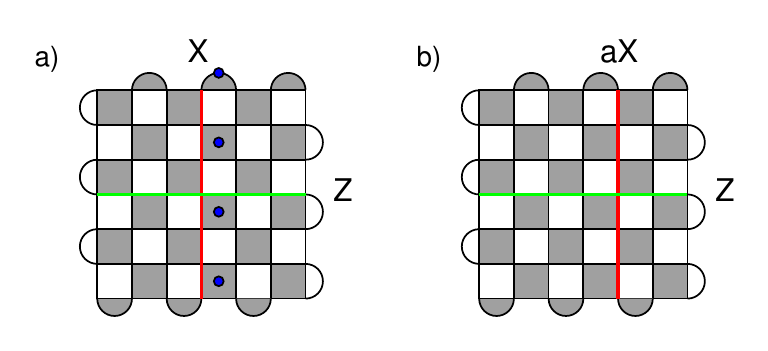}}
\end{center}
\caption{a) Before movement. The product of the $\pm$1 measurement results of the stabilizers marked by blue circles shall be denoted $a$. b) After movement, the new logical operator is related to the old by $a$.}\label{operator_movement}
\end{figure}

Logical $\ket{0}$/$\ket{+}$, namely $\ket{0_L}$/$\ket{+_L}$, can be prepared as shown in Fig.~\ref{initialization}. After preparation, logical operators can be moved around using products of stabilizers, for example as in Fig.~\ref{operator_movement}. Logical $Z$ or $X$ measurements can be achieved by measuring each data qubit in the $Z$ or $X$ basis and, after correction, taking the product of these $\pm$1 values along the indicated lines, with final modification by any sign accumulated through operator movement or manipulation to obtain the actual result.

\section{XX and ZZ logical measurement}

\begin{figure}
\begin{center}
\resizebox{75mm}{!}{\includegraphics{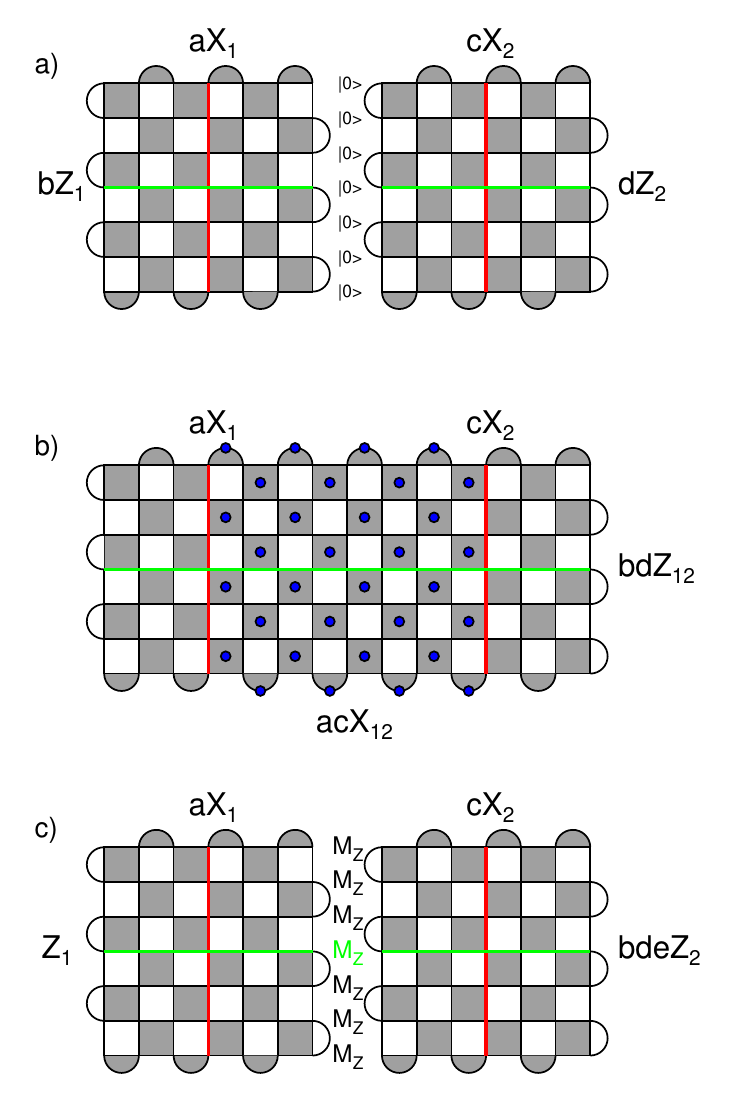}}
\end{center}
\caption{a) Getting ready to measure the logical operator $X_{12}=X_1X_2$. b) The product of the stabilizers marked by blue circles gives us the eigenvalue of the tensor product of $X$ operators along the red lines. This must then be modified by the current signs of tensor products of operators along these lines to give the actual desired result. This pattern of stabilizers is measured $d$ times. c) After splitting, the eigenvalue of the measurement in green is denoted by $e$, and this and the sign of the $Z_{12}$ operator can be associated with either $Z_1$ or $Z_2$.}\label{XX}
\end{figure}

\begin{figure*}
\begin{center}
\resizebox{150mm}{!}{\includegraphics{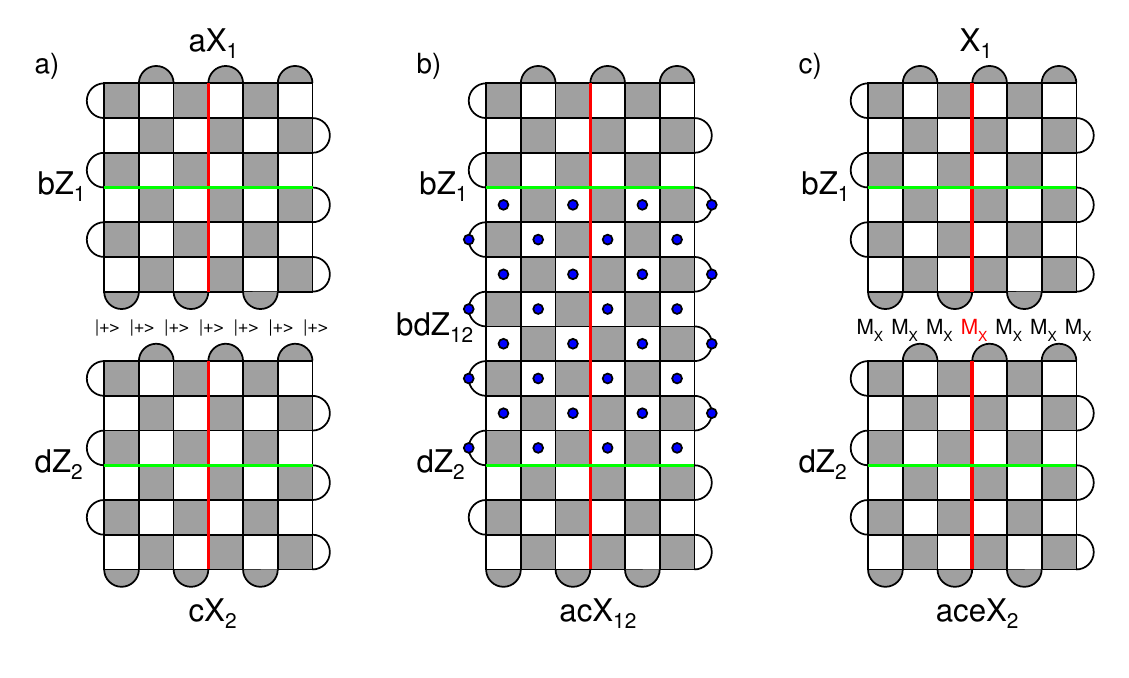}}
\end{center}
\caption{a) Getting ready to measure the logical operator $Z_{12}=Z_1Z_2$. b) The product of the stabilizers marked by blue circles gives us the eigenvalue of the tensor product of $Z$ operators along the green lines. This must then be modified by the current signs of tensor products of operators along these lines to give the actual desired result. This pattern of stabilizers is measured $d$ times. c) After splitting, the eigenvalue of the measurement in red is denoted by $e$, and this and the sign of the $X_{12}$ operator can be associated with either $X_1$ or $X_2$.}\label{ZZ}
\end{figure*}

\begin{figure}
\begin{center}
\resizebox{57mm}{!}{\includegraphics{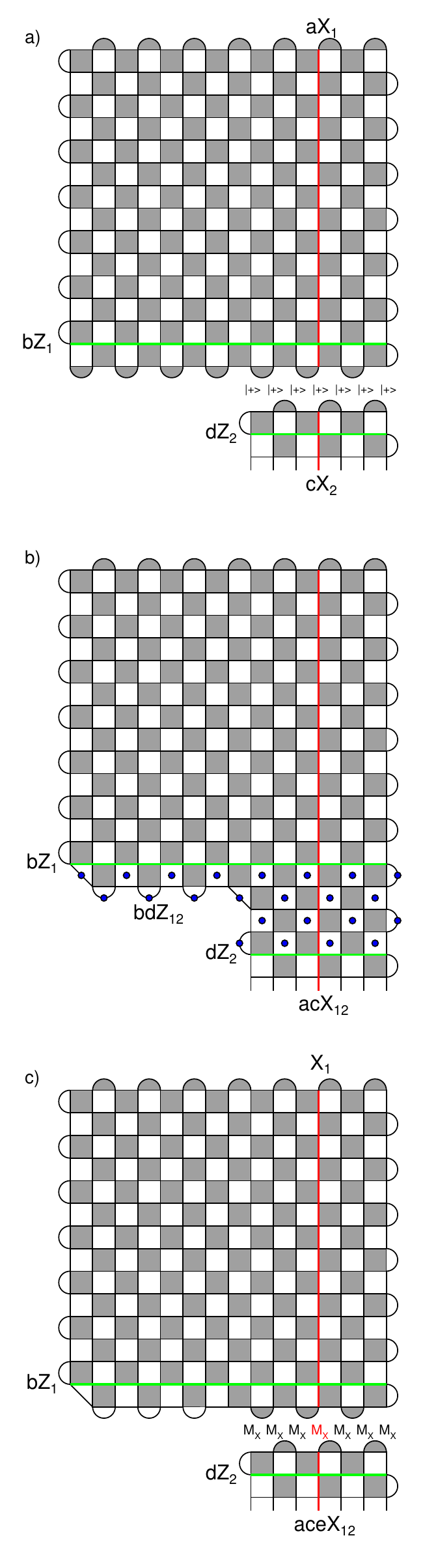}}
\end{center}
\caption{Analogue of Fig.~\ref{ZZ}, namely logical $ZZ$ measurement, given qubits of unequal size.}\label{ZZ_big_small}
\end{figure}

Procedures for measuring logical $XX$ and $ZZ$ given qubits of equal size are shown in Figs.~\ref{XX}--\ref{ZZ}. All described actions with all stabilizer and individual qubit measurements occur only post error correction. It is also useful to know how to measure such operators given qubits of unequal size, and an example of an unequal size $ZZ$ logical measurement can be found in Fig.~\ref{ZZ_big_small}.

\section{Multi-body operator measurements}

\begin{figure}
\begin{center}
\resizebox{75mm}{!}{\includegraphics{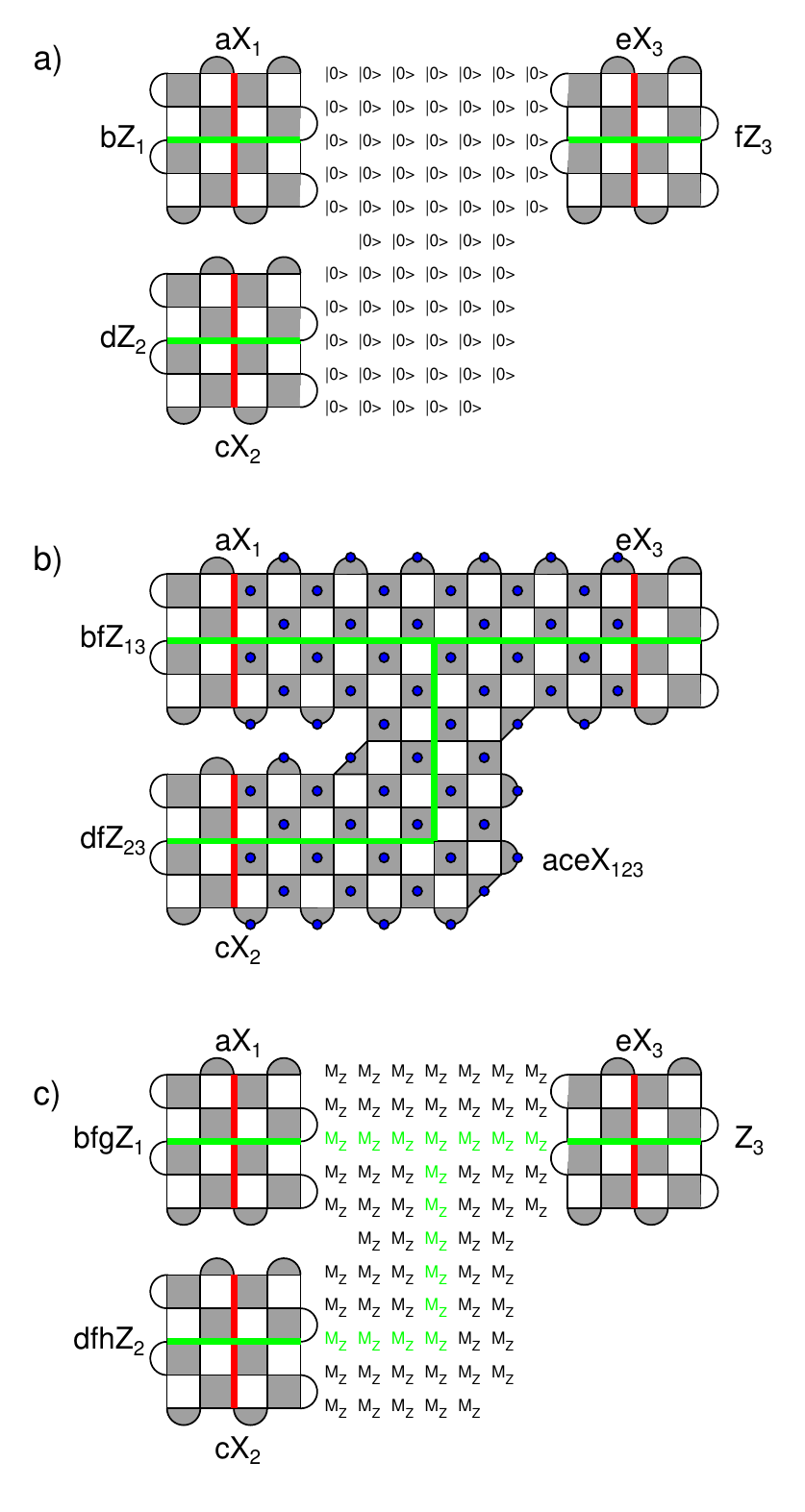}}
\end{center}
\caption{a) Getting ready to measure the logical operator $X_{123}=X_1X_2X_3$. b) The product of the stabilizers marked by blue circles gives us the eigenvalue of the tensor product of $X$ operators along the red lines. This must then be modified by the current signs of tensor products of operators along these lines to give the actual desired result. This pattern of stabilizers is measured $d$ times. c) After splitting, the eigenvalue of the chain of measurements in green connecting $Z_1$ to $Z_3$ is denoted by $g$, and the chain connecting $Z_2$ to $Z_3$ by $h$. The signs of the $Z_{13}$ and $Z_{23}$ operators are most conveniently associated with $Z_1$ and $Z_2$.}\label{multi_body_X}
\end{figure}

\begin{figure}
\begin{center}
\resizebox{75mm}{!}{\includegraphics{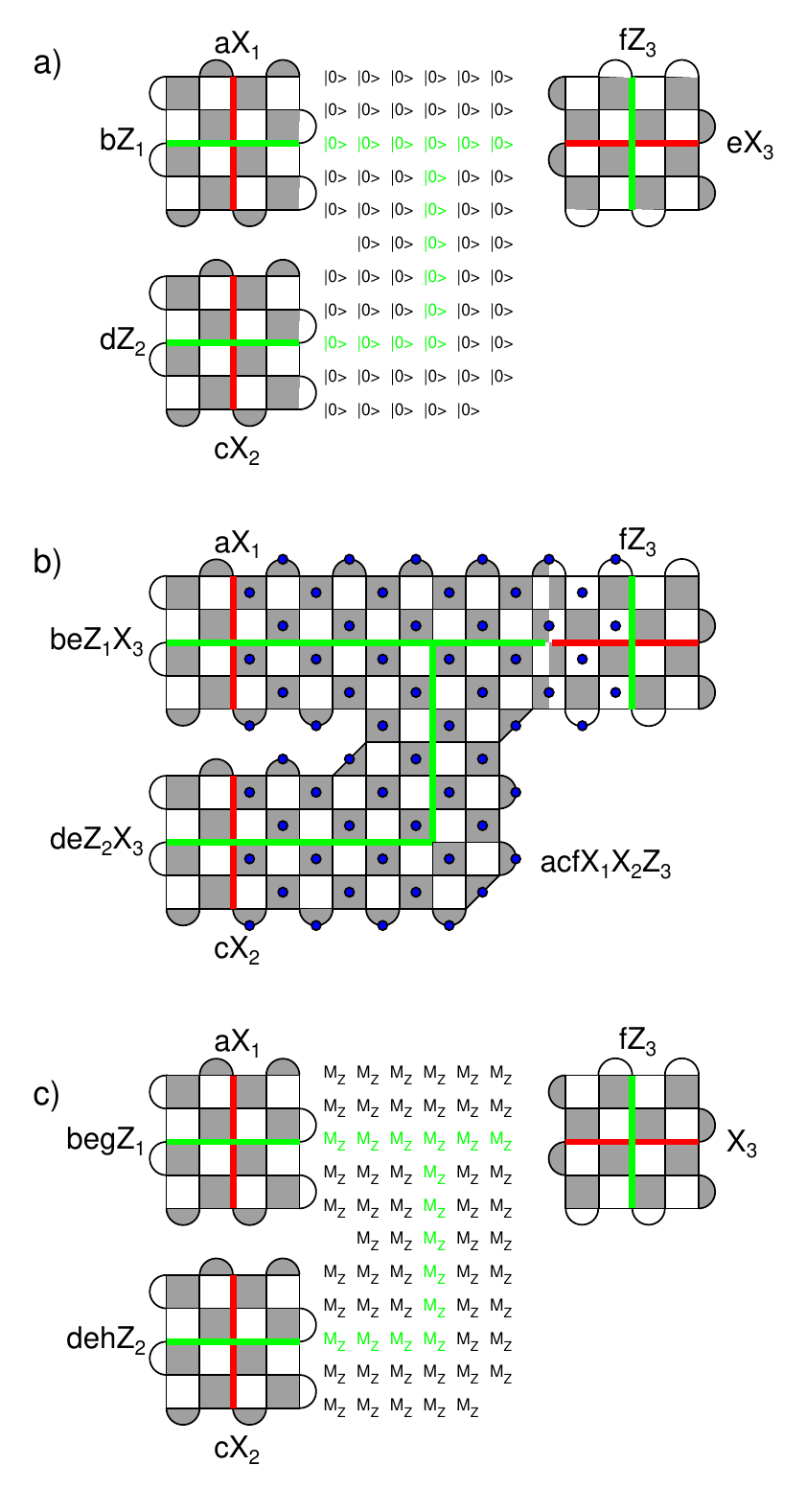}}
\end{center}
\caption{a) Getting ready to measure the logical operator $X_1X_2Z_3$. b) The product of the stabilizers marked by blue circles gives us the eigenvalue of the tensor product $X_1X_2Z_3$. This must then be modified by the current signs of these individul operators to get the actual desired result. This pattern of stabilizers is measured $d$ times. c) After splitting, the eigenvalue of the chain of measurements in green connecting $Z_1$ to $X_3$ is denoted by $g$, and the chain connecting $Z_2$ to $X_3$ by $h$. The signs of the $Z_1X_3$ and $Z_2X_3$ operators are most conveniently associated with $Z_1$ and $Z_2$.}\label{multi_body_XXZ}
\end{figure}

A powerful feature of lattice surgery is the ability to compactly measure multi-body $X$ and $Z$ operators. The explicit steps to perform this for a multi-body $X$ measurement are shown in Fig.~\ref{multi_body_X}. Fig.~\ref{multi_body_XXZ} shows the procedure for logical $XXZ$. When measuring multi-body mixed $X$ and $Z$ operators, it can be necessary to rotate logical qubits.

\section{State injection}

\begin{figure}
\begin{center}
\resizebox{55mm}{!}{\includegraphics{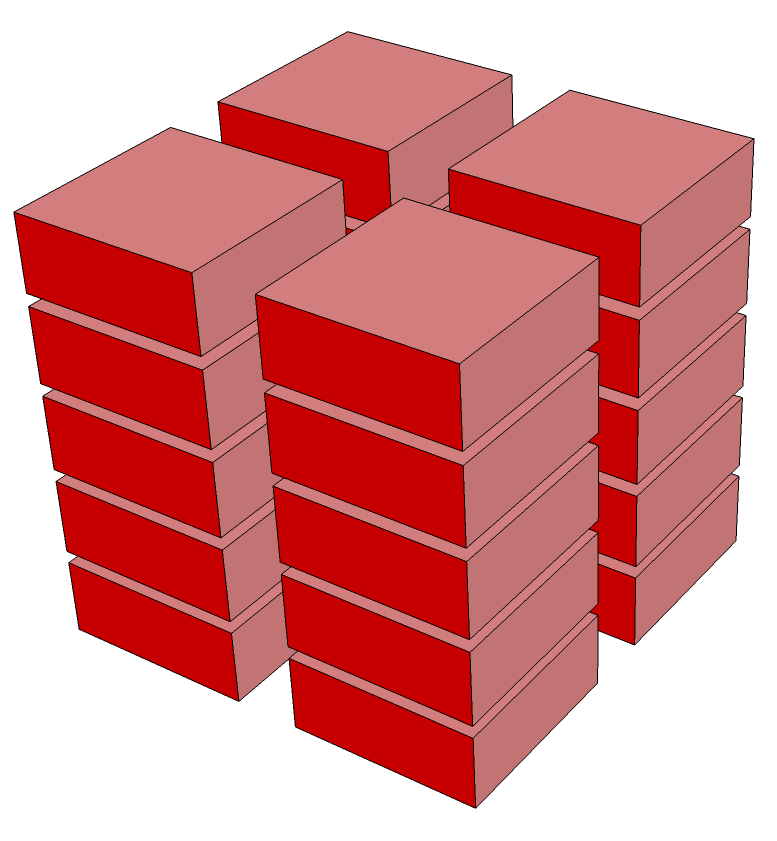}}
\end{center}
\caption{In this figure, and all subsequent 3D figures, time runs vertically. Each red cuboid is an attempt to inject the state $\ket{T}=(\ket{0}+e^{i\pi/4}\ket{1})/\sqrt{2}$ into a distance 7 code. In 15 rounds of a distance 15 code, there is space for 20 injection attempts. When an injection is successful, which occurs 50\% of the time \cite{Li15}, no further injection attempts are made and the injected logical qubit is expanded to distance 15 (Fig.~\ref{expand}) for more reliable storage.}\label{injection_cluster}
\end{figure}

\begin{figure}
\begin{center}
\resizebox{75mm}{!}{\includegraphics{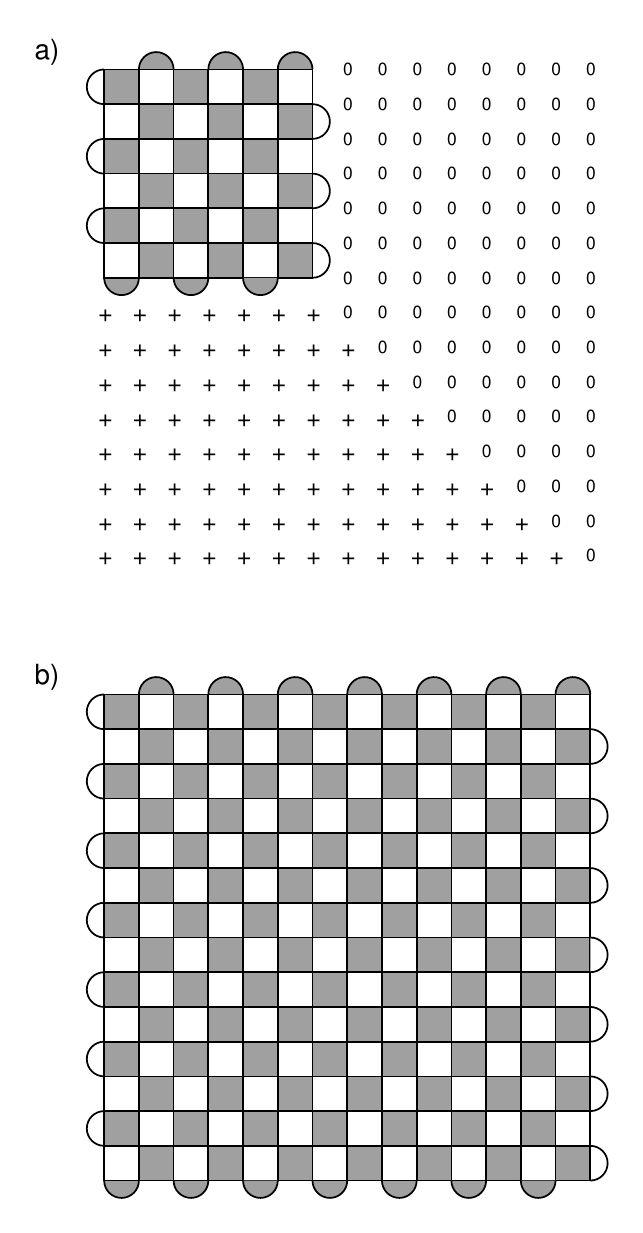}}
\end{center}
\caption{a) Getting ready to expand a distance 7 logical qubit to distance 15. Initialization to $\ket{0}$ and $\ket{+}$ is denoted by 0 and + for visual simplicity. b) After initialization, the complete pattern of stabilizers is measured. The full benefit of the increased code size will not be reached until the complete pattern has been measured 15 times ensuring that the new stabilizers are known with distance 15 confidence, however the single-round logical error rate will be strictly lower even during the first complete error detection round.}\label{expand}
\end{figure}

The output probability of error $p_o$ of 15-1 state distillation \cite{Brav05,Reic05} depends on the injected state error probability $p_i$ as $p_o=35p_i^3$. Using the state injection techniques of \cite{Li15}, it is seems possible with approximately 50\% heralded chance of success to create a distance 7 rotated logical qubit in an arbitrary state with probability of error on acceptance $p_i$ equal to the physical 2-qubit gate error rate $p_2$, however note that the simulations in \cite{Li15} were performed using un-rotated square logical qubits. We are interested in injecting the state $\ket{T}=(\ket{0}+e^{i\pi/4}\ket{1})/\sqrt{2}$. The injection process itself takes two rounds of surface code error detection. We assume it takes an amount of time equal to one further round to determine whether the injection was successful. We shall base our construction on clusters of 20 injection attempts (Fig.~\ref{injection_cluster}), ensuring a heralded probability of failure of each cluster of order $10^{-6}$. When a state is injected successfully, it is expanded to distance 15 (Fig.~\ref{expand}).

\section{The S/S$^\dag$ gate}

We will choose to implement these gates using the method shown in Fig.~7 of \cite{Brow16}. We will assume that given a $d=15$ surface code it is possible in 15 error detection cycles to implement S/S$^\dag$ with distance no less that 7. Note that implementing this gate in this manner typically involves measuring some stabilizers of extended size, irregular shape, and increased weight.

\section{The T/T$^\dag$ gate}

\begin{figure}
\begin{center}
\resizebox{75mm}{!}{\includegraphics{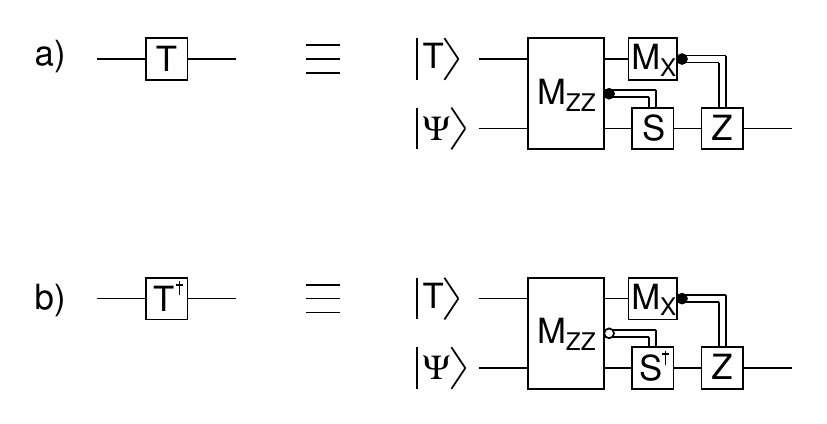}}
\end{center}
\caption{a) T gate and b) T$^\dag$ built out of lattice-surgery structures. Solid dots represent classically controlled execution if the associated measurement is 1 (-1 eigenstate), the open dot is for 0 (+1 eigenstate) triggered execution.}\label{T}
\end{figure}

Fig.~\ref{T} shows how this can be build out of circuit elements already described.

\section{Half distance rotation}

\begin{figure}
\begin{center}
\resizebox{75mm}{!}{\includegraphics{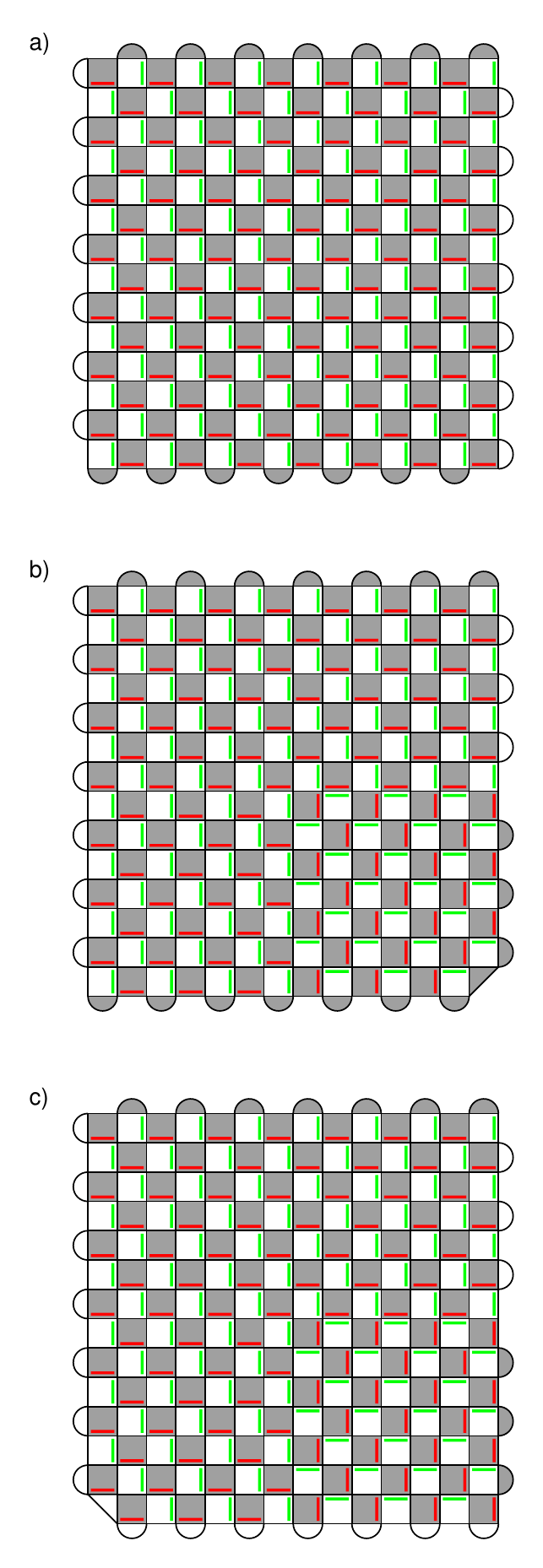}}
\end{center}
\caption{a) Initial distance 15 surface code marked up to show the last 2 qubits interacted with when measuring each stabilizer. b) Modification of the boundary stabilizers and interaction pattern, maintained for 7 rounds. c) Further boundary modification leading to logical Z being supported on the lower half of the right boundary, maintained for 7 rounds. This is then followed by 7 rounds of b), then back to the a) pattern.}\label{rotate_d15}
\end{figure}

\begin{figure}
\begin{center}
\resizebox{25mm}{!}{\includegraphics{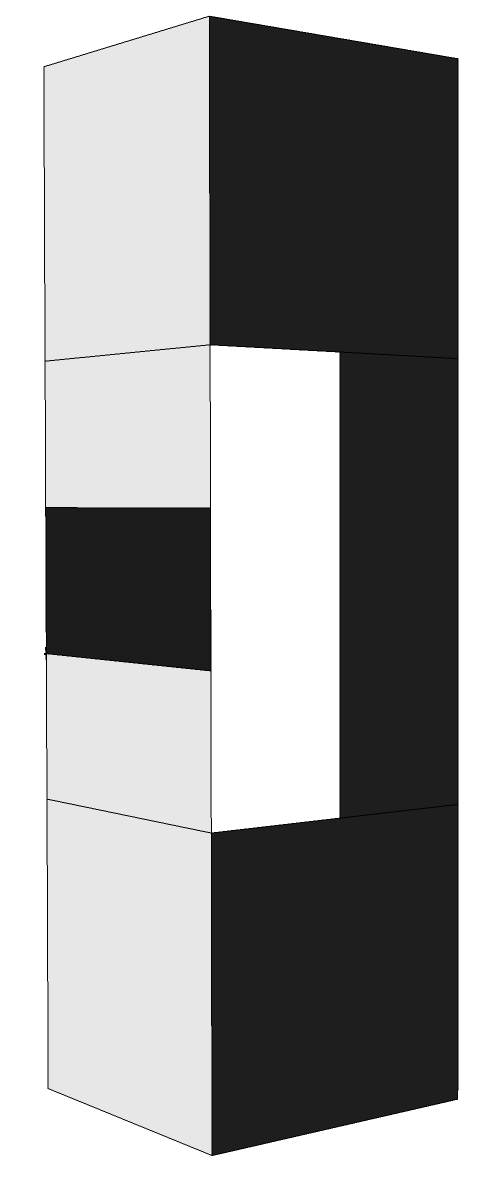}}
\end{center}
\caption{3D version of Fig.~\ref{rotate_d15} showing an additional 15 rounds of standard distance 15 both before and after the half rotation. Light shading represents where logical Z is supported, dark shading represents where logical X is supported.}\label{half_rotate}
\end{figure}

It is frequently useful to be able to rotate square logical qubits by 90$^\circ$. When space is limited and this rotation must be done in place, the code distance will in general be reduced. In Figs.~\ref{rotate_d15}--\ref{half_rotate}, we describe a simple rotation method that halves the code distance but will be sufficient for our purposes.

\section{Distillation}

\begin{figure*}
\begin{center}
\resizebox{175mm}{!}{\includegraphics{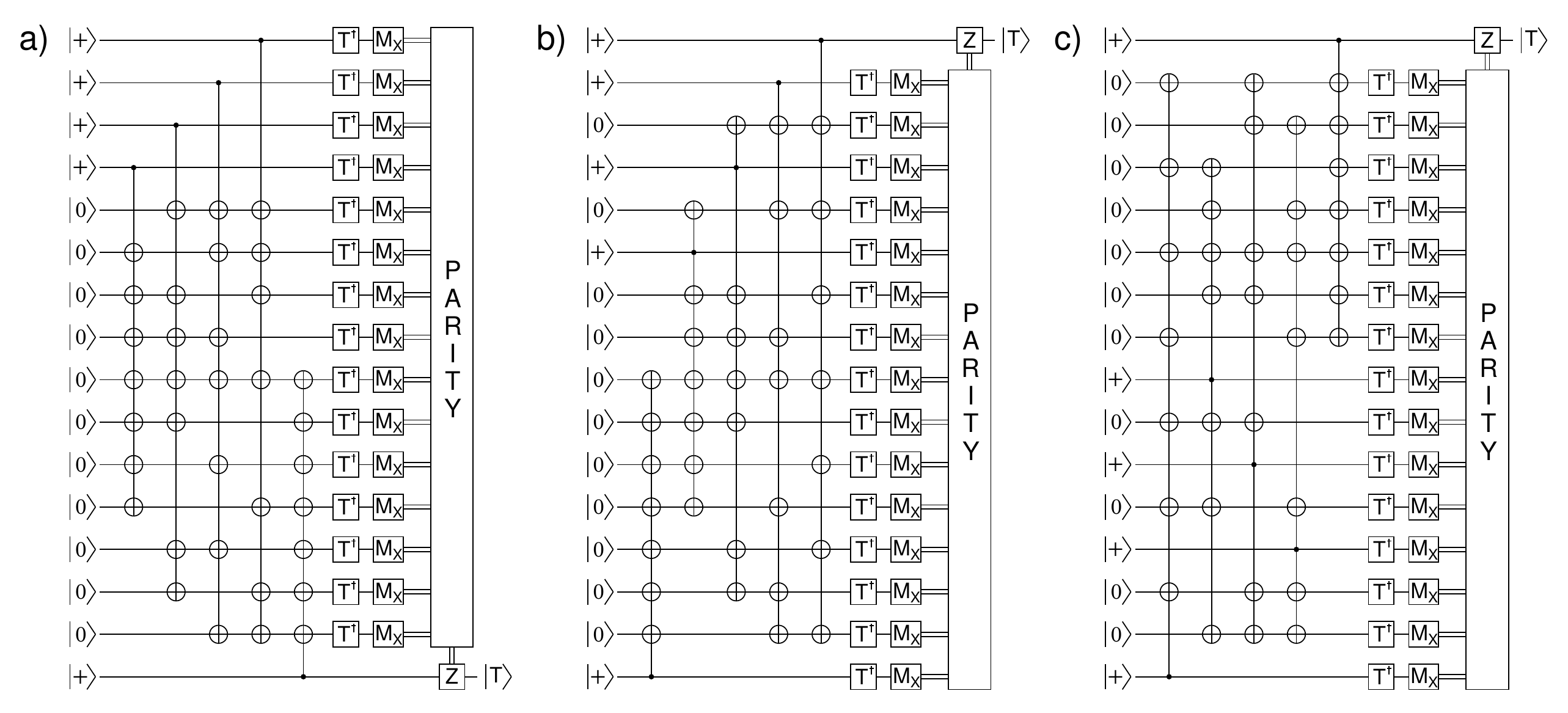}}
\end{center}
\caption{Three versions of the state distillation circuit, identical up to qubit permutation.}\label{distill_stabilizers_all}
\end{figure*}

Consider Fig.~\ref{distill_stabilizers_all}a. The first 4 columns of CNOTs prepare a superposition of computational basis states, each of which has either 8 or no 1s in it. Looked at another way, each column of CNOTs modifies the initial state such that it has an additional stabilizer generator corresponding to X on the control and every target.

These 8-body X stabilizer generators are constructed such that the various patterns of Xs that touch the top 15 qubits uniquely correspond to every binary number from 1 to 15. This enables a single Z error after state preparation to be located, and arbitrary pairs of Z errors to be detected. This state can be considered logical $\ket{0}$ of a distance 3 code.

The 5th and final column of CNOTs corresponds to the controlled application of logical X, meaning the superposition $(\ket{0_L}+\ket{1_L})/\sqrt{2}$ is prepared. To be more precise, given the 16th qubit, $(\ket{0_L}\ket{0}+\ket{1_L}\ket{1})/\sqrt{2}$ is prepared. States in $\ket{1_L}$ contain 15 or 7 1s.

The number of 1s in $\ket{0_L}$ and $\ket{1_L}$ implies that applying transversal T$^\dag$ will result in $(\ket{0_L}\ket{0}+e^{i\pi/4}\ket{1_L}\ket{1})/\sqrt{2}$. Z errors during the T$^\dag$ gates can then be detected by measuring each qubit in the X basis, as the eigenvalues of the 4 X stabilizer generators can be reconstructed by taking the appropriate product of X measurements. If any generator has a negative eigenvalue, the 16th qubit is discarded.

If the probability of a Z error during T$^\dag$ is $p$, the probability of rejection is approximately 15$p$. The collective effect of the X measurements is a logical X measurement, which cuts the output down to $(\ket{T}=(\ket{0}+e^{i\pi/4}\ket{1})/\sqrt{2}$ up to a possible Z correction. 35 combinations of 3 Z errors are undetectable and lead to an erroneous output of $Z\ket{T}$, hence the input-output error relationship $p_o=35p_i^3$ already mentioned.

\begin{figure}
\begin{center}
\resizebox{75mm}{!}{\includegraphics{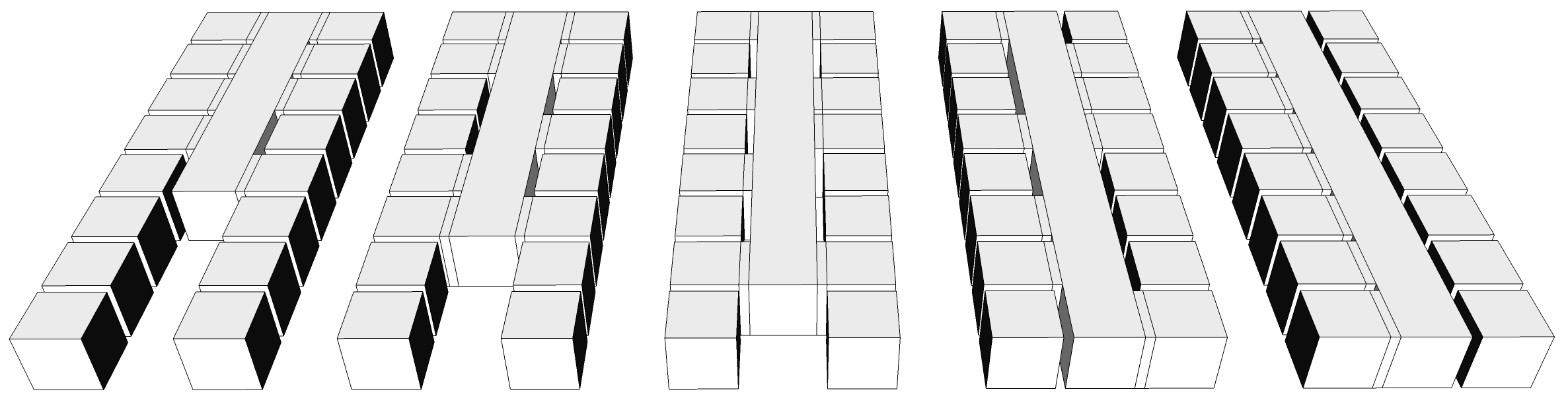}}
\end{center}
\caption{5 structures corresponding to the 5 columns of CNOTs in Fig.~\ref{distill_stabilizers_all}a. Instead of implementing CNOTs, the same state can be prepared with multi-body X measurements. In each structure the left 8 blocks correspond to the lower 8 qubits, and the right 8 blocks to the upper 8 qubits (in reverse order so the top qubit is the bottom right block). The element in the center of each structure performs the appropriate multi-body X measurement.}\label{lattice_surgery_layers}
\end{figure}

The 5th and final column of CNOTs is structurally identical to the first 4 columns, meaning it can be considered the 5th stabilizer generator of a 16 qubit code. This suggests an efficient method of preparing this state using lattice surgery, namely the direct measurement of the 5 stabilizer generators. Fig.~\ref{lattice_surgery_layers} explicitly illustrates such measurements.

\begin{figure}
\begin{center}
\resizebox{75mm}{!}{\includegraphics{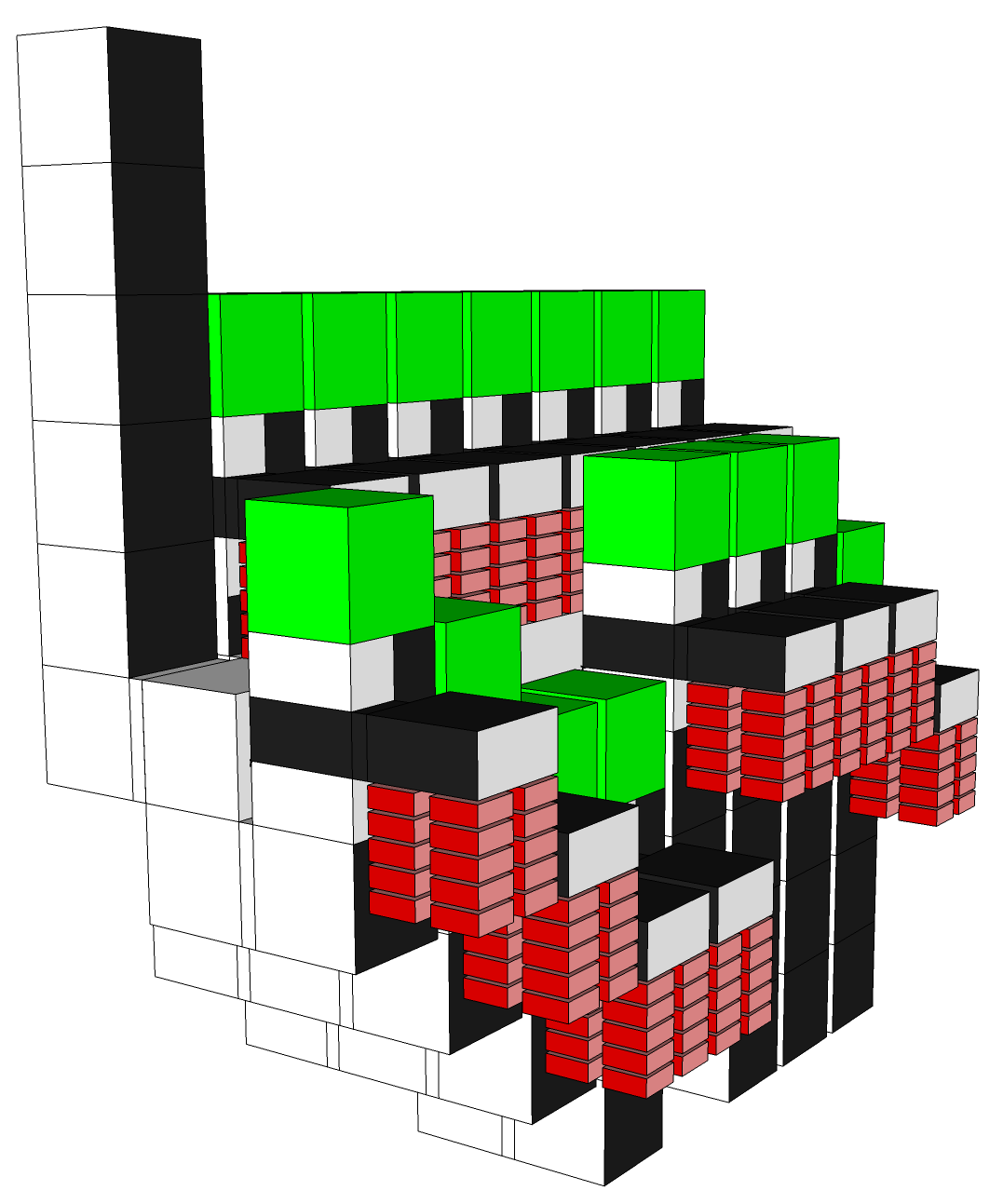}}
\end{center}
\caption{The first round of state distillation begins with 5 X stabilizer generator measurements taken from Fig.~\ref{lattice_surgery_layers} and stacked one atop another, with unnecessary blocks deleted both from top and bottom (no need to create a logical qubit before its first interaction, or hold onto it after its last). As soon as possible, appropriate outputs are half-distance rotated and unequal-size ZZ measurement interacted with the expanded output of multi-attempt state injection. The green box capping this is an S$^\dag$ gate that is included or not based on whether the ZZ measurement yields 0 or 1 (Fig.~\ref{T}b).}\label{L1}
\end{figure}

Putting it all together, after preparing the 16 qubit state, 15 of these qubits get half-distance rotated and ZZ measured with the expanded output of multi-attempt state injection as shown in Fig.~\ref{L1}.

If we assume that the probability of a Z error during the complete T$^\dag$ structure is $p_i=10^{-3}$, the output probability of error will be $p_o=3.5\times10^{-8}$. This might be sufficient for medium-term algorithms, however further error suppression is likely necessary for long-term algorithms.

\begin{figure}
\begin{center}
\resizebox{75mm}{!}{\includegraphics{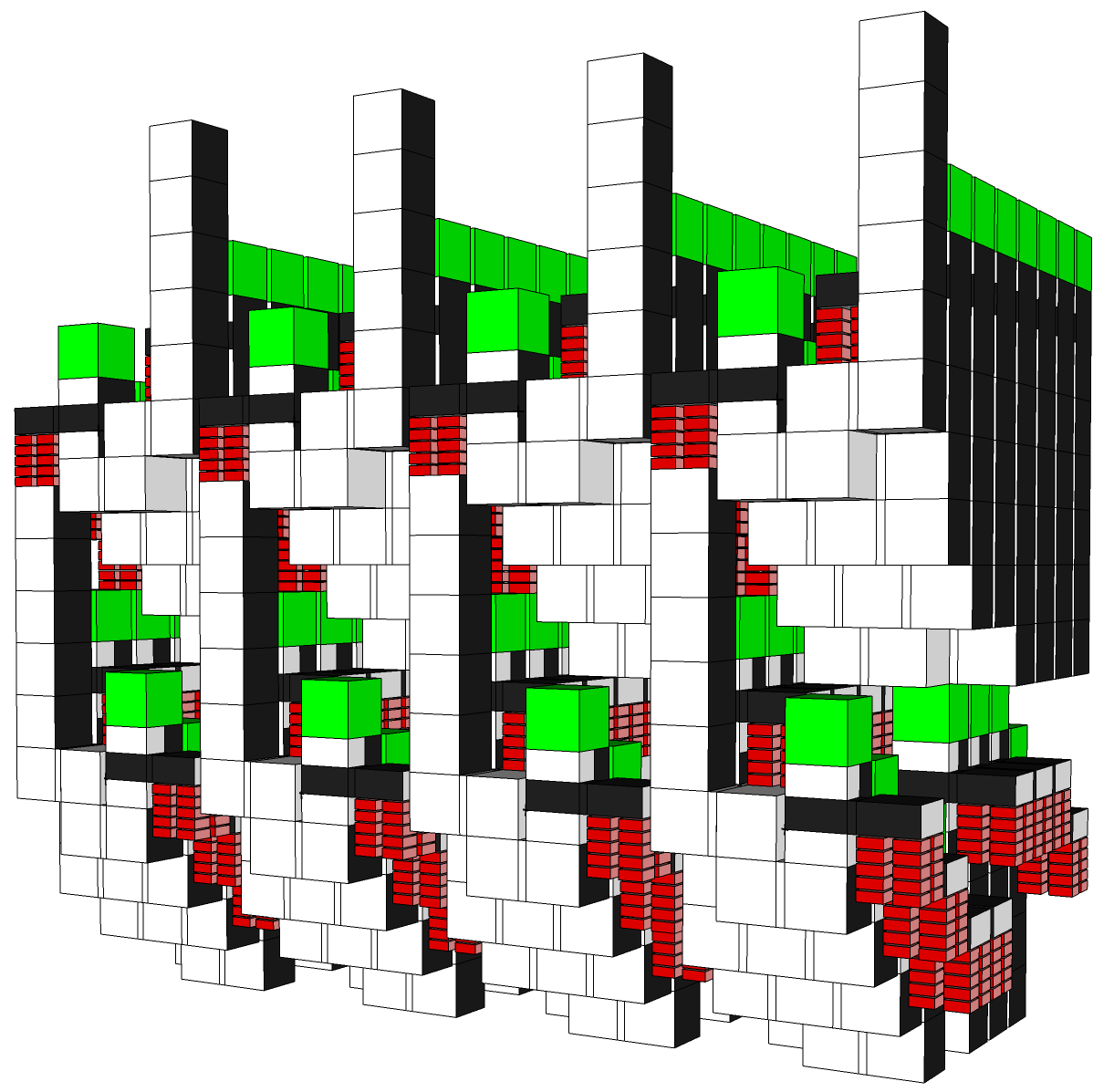}}
\end{center}
\caption{8 attempts to produce a $\ket{T}$ state. Note that the 2nd layer of distillation is mirror reflected from the first.}\label{L1_rack}
\end{figure}

If we wish to repeat state distillation, we need to attempt to prepare at least 16 first-level $\ket{T}$ states to reliably get 15 inputs given the 15$p_i$ failure rate. 8 packed level one distillations are shown in Fig.~\ref{L1_rack}. The effective height of each layer of the structure is 6.5$d$, where $d$ is explicitly fixed at 15.

\begin{figure}
\begin{center}
\resizebox{75mm}{!}{\includegraphics{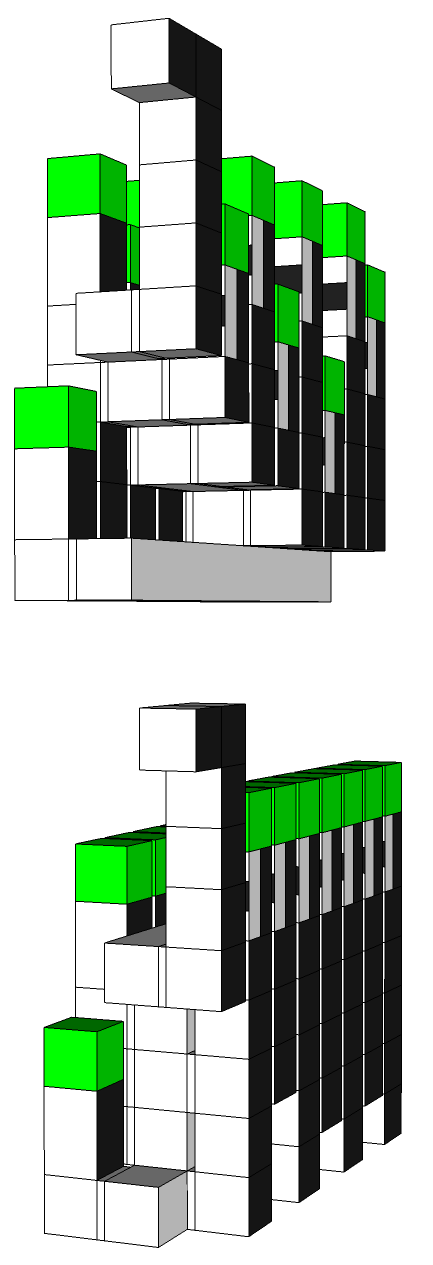}}
\end{center}
\caption{Top: lattice surgery version of Fig.\ref{T}b). Bottom: lattice surgery version of Fig.\ref{T}c). Block position convention as in Fig.~\ref{L1}.}\label{L2_isolated}
\end{figure}

\begin{figure*}
\begin{center}
\resizebox{85mm}{!}{\includegraphics{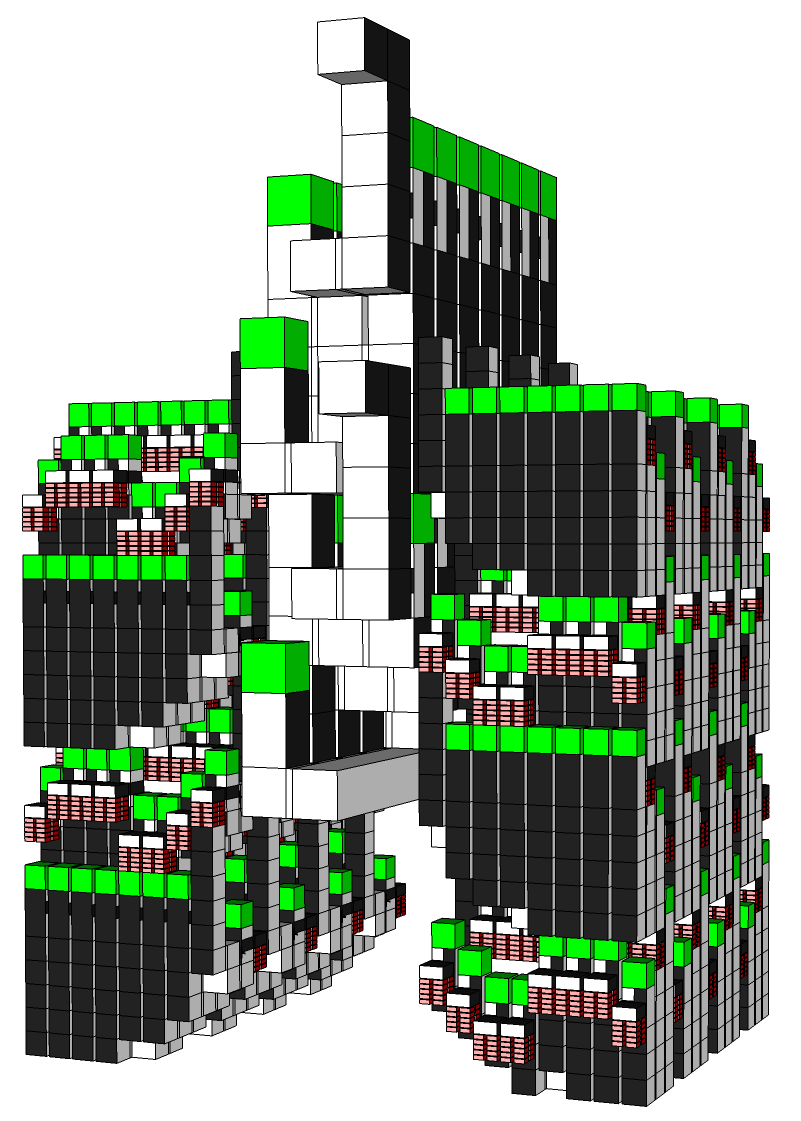}}
\end{center}
\caption{Two rounds of two levels of state distillation.}\label{new_two_levels2}
\end{figure*}

For the second level of distillation, we will use the two circuits of Fig.\ref{T}b)--c). These circuits have been designed such that the last round of CNOTs of b) slots into the first round of CNOTs of c), and the last round of CNOTS of c) slots into the first round of CNOTs of b). This property ensures that the lattice surgery versions of these circuits (Fig.~\ref{L2_isolated}) fit snuggly together, reducing the overall height and therefore time of execution (Fig.~\ref{new_two_levels2}).

\section{Hadamard}

The logical Hadamard gate, being a Clifford gate, can in principle be performed entirely in classical processing, by simply relabeling the logical X and Z operators \cite{Liti18}. When using lattice surgery, this comes at the cost of occasionally needing to double the size of the logical qubit, and being able to measure logical Y operators, which doubles the size again. Furthermore, physical-level stabilizer measurements of weight greater than 4 and structure that may or may not be suitable for the underlying hardware are required.

\begin{figure*}
\begin{center}
\resizebox{170mm}{!}{\includegraphics{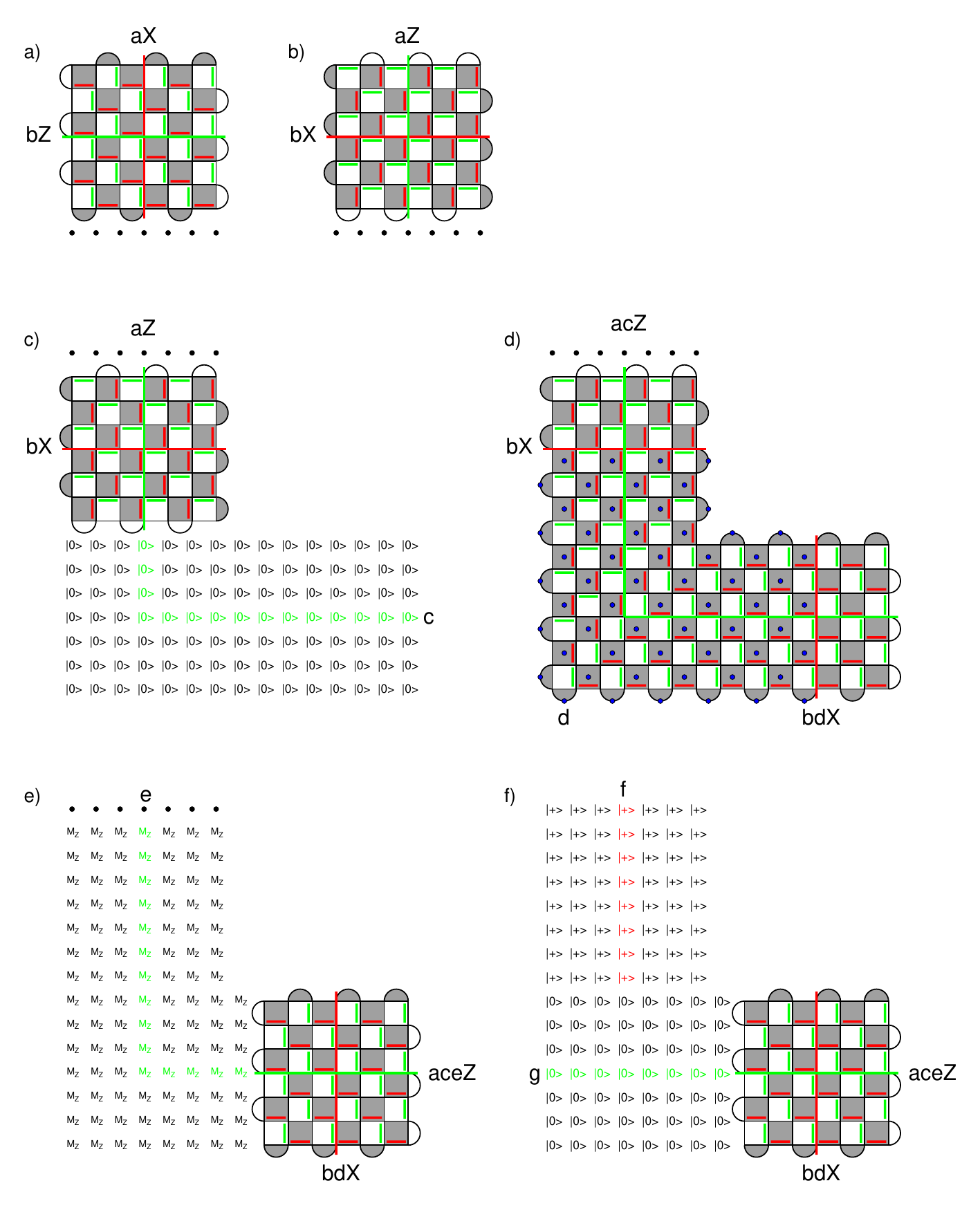}}
\end{center}
\caption{a) Initial state. b) State after transversal Hadamard. c) State after swap downwards and preparing to expand the logical qubit. d) Expanded logical qubit. Note the orientation of circuits around the corner to ensure the full code distance is preserved. Blue dots show how to move the logical X operator using the product d of stabilizers. e) Contraction of the logical qubit. f) Initialization of the way back to the original position. This and subsequent steps could be omitted if a return to the original position is not required.}\label{Hadamard}
\end{figure*}

\begin{figure*}
\begin{center}
\resizebox{175mm}{!}{\includegraphics{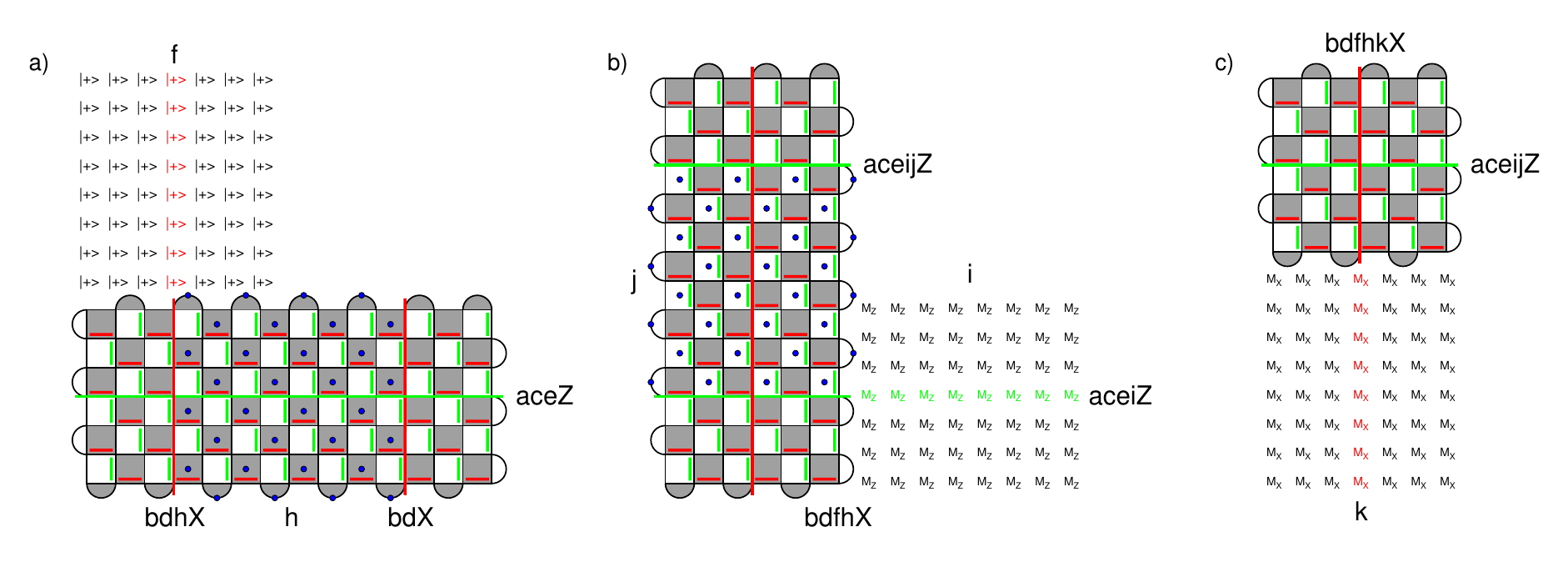}}
\end{center}
\caption{a) Stabilizer pattern to move logical qubit back to original position. Movement of logical X operator. b) Trimming and movement of logical Z operator. c) Trimming and movement of logical X operator.}\label{Hadamard2}
\end{figure*}

We are therefore motivated to present a method of performing Hadamard that does not change the definitions of the logical operators and uses simple underlying circuits. Our proposal can be found in Figs.~\ref{Hadamard}--\ref{Hadamard2}.

\section{CNOT}

\begin{figure}
\begin{center}
\resizebox{75mm}{!}{\includegraphics{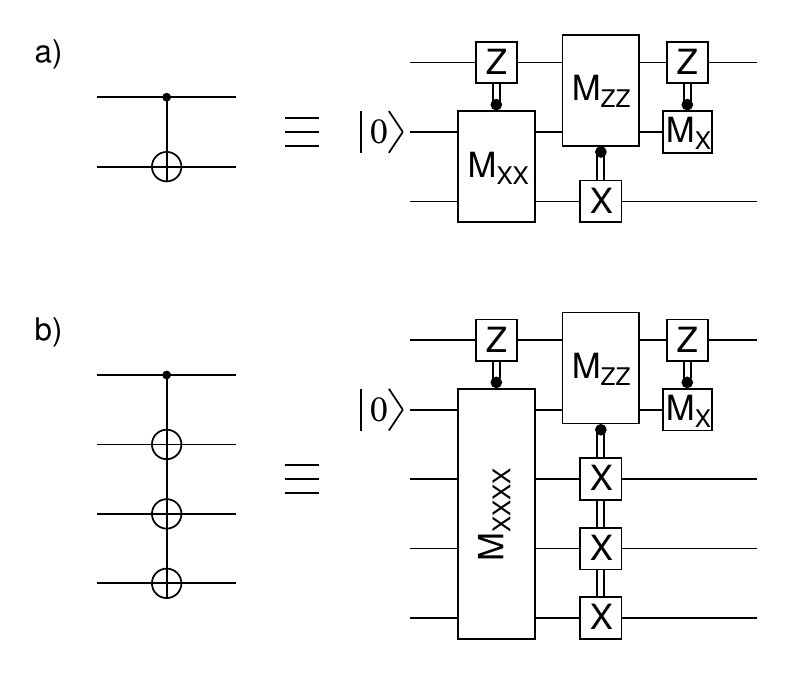}}
\end{center}
\caption{a) Lattice surgery implementation of CNOT. b) Extension to single-control multiple-target CNOT.}\label{CNOT}
\end{figure}

CNOT is an extremely common gate in many quantum algorithms. Regular CNOT and single-control multiple-target CNOT can be implemented in similar manners (Fig.~\ref{CNOT}). Note that the total time required is 2$d$ as the ancilla initialization and measurement can both be performed transversely.

\section{CZ}

\begin{figure}
\begin{center}
\resizebox{75mm}{!}{\includegraphics{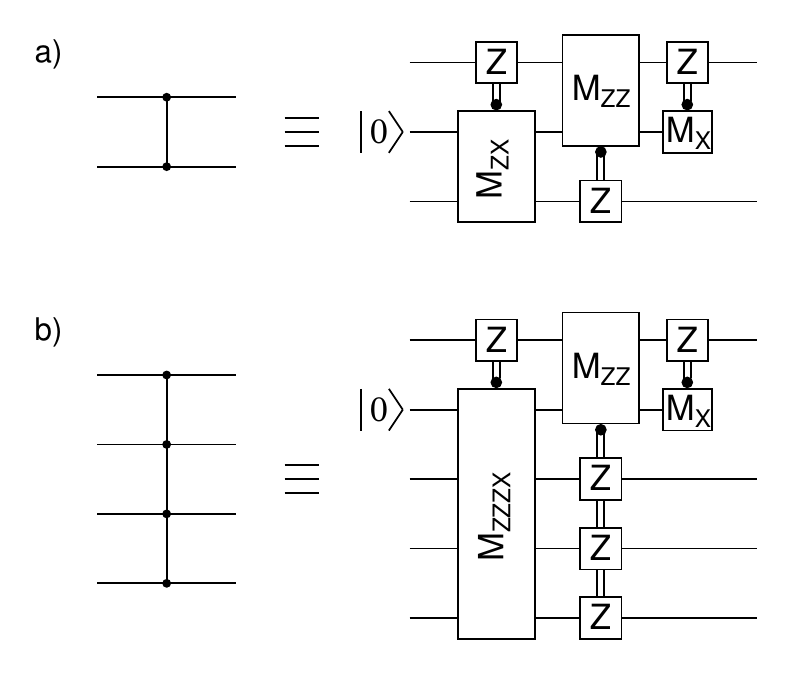}}
\end{center}
\caption{a) Lattice surgery implementation of CZ. b) Extension to single-control multiple-target CZ.}\label{CZ}
\end{figure}

The CNOT of the previous section can be modified to create a CZ gate simply by changing one of the operators being measured (Fig.~\ref{CZ}).

\section{Swap}

\begin{figure}
\begin{center}
\resizebox{75mm}{!}{\includegraphics{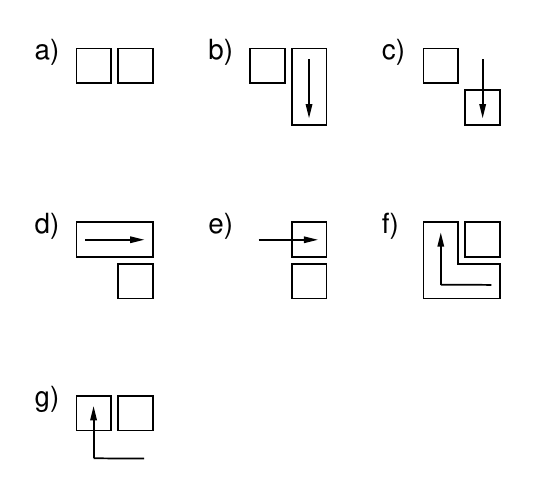}}
\end{center}
\caption{a) Initial configuration. b)--c) Downward movement, $d$ error detection rounds. d)--e) Rightward movement, $d$ error detection rounds. f)--g) L-shaped return movement using the techniques in Fig.~\ref{Hadamard2}, a final $d$ error detection rounds for a total of 3$d$.}\label{swap}
\end{figure}

Logical qubits can be moved around one another using a series of single logical qubit moves (Fig.~\ref{swap}).

\section{Algorithm overhead}

\begin{figure}
\begin{center}
\resizebox{75mm}{!}{\includegraphics{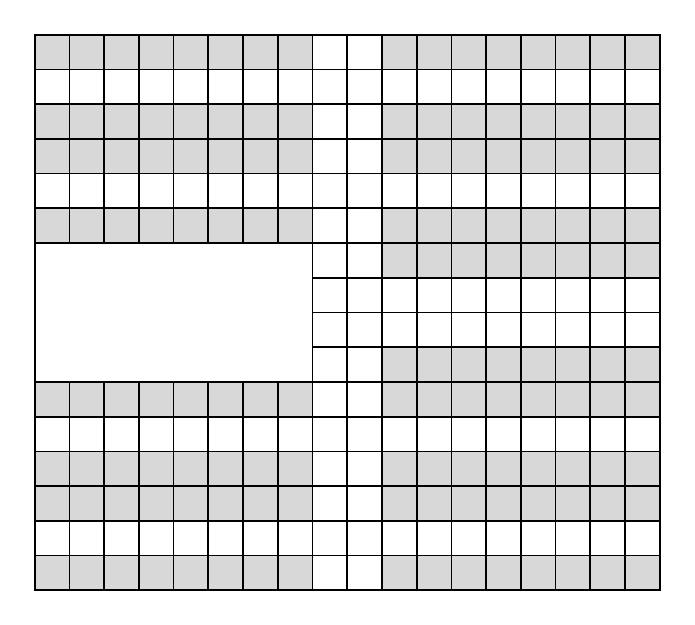}}
\end{center}
\caption{Example layout of an algorithm showing a single level of distillation (large open rectangle), data logical qubits (shaded squares), and ancilla logical qubits for communication and interaction (white squares).}\label{algorithm}
\end{figure}

In keeping with \cite{Babb18}, we shall focus on the situation where you only distill one $\ket{T}$ state at a time (Fig.~\ref{algorithm}). To ensure a high chance of keeping pace with a superconducting quantum computer, we shall process surface code detection events using the simple techniques of \cite{Fowl10} rather than the more computationally expensive and better logical error suppressing techniques of \cite{Fowl13g}. This leads to an approximate logical error rate per round of error detection of $p_L=0.1(100p)^{(d+1)/2}$.

Calculating the overhead of a Clifford+T algorithm is then straightforward, if you can approximate the algorithm as being execution-time dominated by the preparation of $\ket{T}$ states, meaning few sections of many Clifford gates, meaning the Clifford gates in the algorithm can be performed in parallel with the $\ket{T}$ preparations. We include a spreadsheet both describing in detail and performing the necessary calculations.

\section{Conclusion}

Using the defect and braiding overhead estimation techniques of \cite{Babb18} an algorithm involving $10^8$ T gates and 100 logical qubits, on hardware with a characteristic gate error rate of $10^{-3}$ and surface code error detection circuit time of 1$\mu$s takes 4.5 hours and requires $1.8\times 10^6$ physical qubits. By contrast, using lattice surgery techniques as described in this paper and the spreadsheet, the same algorithm would run in 5.4 hours, so comparable, but require just $3.7\times 10^5$ physical qubits. This near factor of 5 qubit saving while maintaining comparable runtime implies defects and braiding should be deprecated in favor of lattice surgery.

\section{Acknowledgments}

We thank Daniel Litinski for discussions leading to the simplification of stabilizers measured during mixed-operator multi-body logical measurements.

\bibliography{References}

\end{document}